\documentclass[%
 reprint, 
 amsmath,amssymb,
 aps,
]{revtex4-2}

\usepackage{graphicx}
\usepackage{dcolumn}
\usepackage{bm}
\usepackage{theorem}
\usepackage{color}
\usepackage{multirow}
\usepackage{amssymb}
\usepackage{subfigure}
\usepackage{mathrsfs}
\usepackage{booktabs}
\usepackage{multirow}
\usepackage{makecell}
\usepackage{hyperref}

{\theorembodyfont{\normalfont\it}
\theoremheaderfont{\normalfont\bf}
\newtheorem{thm}{ Theorem}
\newtheorem{dfn}[thm]{ Definition}
\newtheorem{lmm}[thm]{ Lemma}

\newtheorem{crl}[thm]{ Corollary}
\newtheorem{asm}[thm]{ Assumption}
\newtheorem{prp}[thm]{ Proposition}
\newtheorem{cjt}[thm]{ Conjecture}
\newtheorem{rmk}[thm]{ Remark}}

{\theorembodyfont{\normalfont}
\theoremheaderfont{\normalfont\it}
\newtheorem{prf}{ Proof:}}

{\theorembodyfont{\normalfont}
\theoremheaderfont{\normalfont\it}
}

{\theorembodyfont{\normalfont}
\theoremheaderfont{\normalfont\it}
}

{\theorembodyfont{\normalfont}
\theoremheaderfont{\normalfont\it}
}

{\theorembodyfont{\normalfont}
\theoremheaderfont{\normalfont\it}
}

{\theorembodyfont{\normalfont}
\theoremheaderfont{\normalfont\it}
}

\newcommand{\bra}[1]{\mbox{$\langle#1|$}}

\newcommand{\ket}[1]{\mbox{$|#1\rangle$}}

\newcommand{\outpro}[2]{\mbox{$\ket{#1}\!\bra{#2}$}}

\newcommand{\proj}[1]{\mbox{$\ket{#1}\!\bra{#1}$}}

\newcommand{\alg}[1]{\begin{align}#1\end{align}}

\newcommand{\nn}{\nonumber}

\newcommand{\mbb}[1]{{\mathbb #1}}

\newcommand{\bthm}[1]{\begin{thm}\label{thm:#1}}
\newcommand{\ethm}{\end{thm}}

\newcommand{\rThm}[1]{Theorem \ref{thm:#1}}
\newcommand{\blmm}[1]{\begin{lmm}\label{lmm:#1}}
\newcommand{\elmm}{\end{lmm}}

\newcommand{\bdfn}[1]{\begin{dfn}\label{dfn:#1}}
\newcommand{\edfn}{\end{dfn}}

\newcommand{\basm}[1]{\begin{asm}\label{asm:#1}}
\newcommand{\easm}{\end{asm}}

\newcommand{\bprp}[1]{\begin{prp}\label{prp:#1}}
\newcommand{\eprp}{\end{prp}}

\newcommand{\bcrl}[1]{\begin{crl}\label{crl:#1}}
\newcommand{\ecrl}{\end{crl}}

\newcommand{\bcjt}[1]{\begin{cjt}\label{cjt:#1}}
\newcommand{\ecjt}{\end{cjt}}

\newcommand{\brmk}[1]{\begin{rmk}\label{rmk:#1}}
\newcommand{\ermk}{\end{rmk}}

\newcommand{\bprf}{\begin{prf}}
\newcommand{\eprf}{\end{prf}}
\newcommand{\laeq}[1]{\label{eq:#1}}
\newcommand{\req}[1]{(\ref{eq:#1})}

\newcommand{\QED}{\hfill$\blacksquare$}

\newcommand{\lapp}[1]{\label{app:#1}}

\newcommand{\rApp}[1]{Appendix \ref{app:#1}}

\newcommand{\bitem}{\begin{itemize}}
\newcommand{\entem}{\end{itemize}}
\newcommand{\benum}{\begin{enumerate}}
\newcommand{\ennum}{\end{enumerate}}

\newcommand{\otm}{\otimes}

\newcommand{\rFig}[1]{Figure \ref{fig:#1}}

\newcommand{\prlsection}[1]{{\it{#1}}.---}

\begin{document}

\preprint{APS/123-QED}

\title{
Gravity-Induced Entanglement of Quantum Clocks\\as a Signature of Genuinely Quantum Local Position Invariance
}
\author{Eyuri Wakakuwa}
\email{e.wakakuwa@gmail.com}
\affiliation{Department of Mathematical Informatics, Nagoya University, Furo-cho, Chikusa-ku, Nagoya, 464-8601, Japan}

\date{\today}

\begin{abstract}
Whether the principles of general relativity extend to a quantum regime remains one of the open questions in modern physics. 
In classical general relativity, Einstein's equivalence principle underpins the interpretation of gravity as spacetime geometry.
However, it is not known whether the principle remains valid when the source of gravity could be quantum. 
Here we show that gravity-induced entanglement (GIE) of quantum clocks provides a framework to characterize local position invariance (LPI) in such a regime, which constitutes one of the subprinciples of the equivalence principle. 
By adapting the existing formulation of quantum LPI to the context of GIE, we analyze two schemes that address complementary aspects of LPI and differ only in the choice of the initial clock state: one in which entanglement will be generated if and only if there is a genuinely quantum violation of LPI, and the other where classical-like LPI violation manifests in the frequency of entanglement oscillation. 
Our results suggest that GIE of quantum clocks offers an approach to investigating fundamental principles of general relativity in the quantum regime, shedding new light on the interplay between quantum mechanics and the theory of gravity.
\end{abstract}

\maketitle


\prlsection{Introduction}
The interplay between quantum mechanics and general relativity remains one of the major challenges in modern physics. 
A central question at this intersection is how gravity behaves when it is sourced by quantum matter.
In classical general relativity, gravity is described as a geometric structure of spacetime.
This geometric picture of gravity is supported by Einstein's equivalence principle \cite{will2018theory,will2014confrontation}.
However, it is not known whether the equivalence principle remains valid when quantum mechanical effects are relevant,
and it is even unclear how it should be formulated in such a regime \cite{aharonov1973quantum,lammerzahl1996equivalence,viola1997testing,davies2004quantum,seveso2017does,zych2017quantum,anastopoulos2018equivalence,geiger2018proposal,hardy2018construction,zych2018quantum,giacomini2020einstein,hardy2020implementation,giacomini2022quantum}.
Consequently, it remains open whether the geometric interpretation of gravity remains valid when the source of gravity is quantum matter.

Zych and Brukner \cite{zych2018quantum,zych2017quantum} proposed a quantum formulation of the equivalence principle by considering quantum mechanical particles in classical background spacetime.
Relying on the mass-energy equivalence,
they formulated the equivalence principle as the equality among the internal Hamiltonians corresponding to the rest mass, the inertial mass and the gravitational mass, respectively.
This approach enables us to treat, in a unified framework, several constituents of the equivalence principle such as the universality of free fall, the local position invariance and the local Lorentz invariance.
Moreover, it uncovers the possibility of ``genuinely quantum'' violation of the equivalence principle, which has no classical counterpart and originates in noncommutativity of the internal Hamiltonians. 
An experimental realization of their approach was also reported in \cite{rosi2017quantum}.
However, their framework is restricted to a classical background spacetime.
It remains unknown whether their framework is applicable to gravity that is sourced by quantum matter, and how it could be tested in such a regime.

\begin{figure}[t]
\includegraphics[bb={-11 0 1179 145}, scale=0.68]{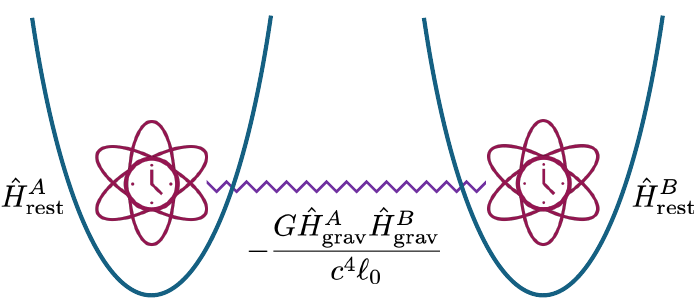}
\caption{Schematic diagram of the proposed setup. Two ``quantum clocks'' are trapped in close proximity by a potential, with the distance $\ell_0$, so that they interact gravitationally. Each clock ticks due to the rest Hamiltonian $\hat{H}_{\rm rest}$, while their interaction is governed by the gravitational Hamiltonian $\hat{H}_{\rm grav}$. The violation of quantum local position invariance, $\hat{H}_{\rm rest}\neq\hat{H}_{\rm grav}$, manifests in the behavior of the entanglement generated between the two clocks.}\label{fig:A}
\end{figure}

In this paper, we propose a setup to investigate local position invariance (LPI) when gravity is sourced by particles that could be in a quantum superposition state.
The setup involves two quantum clocks interacting gravitationally, and uses gravity-induced entanglement (GIE) \cite{bose2017spin,marletto2017gravitationally} generated between them as a signature of the possible violation of LPI.
By extending the framework of Zych and Brukner, we show that the two complementary aspects of LPI violation, namely, ``genuinely quantum'' violation and ``classical-like'' one, induces distinct signatures on the behavior of entanglement.
Importantly, these two aspects can be addressed in a single setup that differ only in the choice of the initial clock states.
Moreover, in one of the schemes, GIE is generated if and only if LPI is violated, hence the generation of GIE itself is a signature of quantum violation of LPI.
This constitutes a crucial advantage over the existing work \cite{bose2023entanglement}, in which QEP violation manifests only quantitatively.
Our result suggests that, though practically challenging, GIE of quantum clocks provides a unified framework for analyzing LPI when gravity is sourced by quantum matter.

Two prior studies are particularly relevant.
Castro-Ruiz et al.~\cite{castro2017entanglement} showed that quantum clocks interacting gravitationally become entangled when initialized in a superposition of internal energies. 
However, their analysis implicitly assumes the quantum equivalence principle to hold, thus excluding the verification of the principle out of their scope.
Bose et al.~\cite{bose2023entanglement} proposed a setup to witness the violation of the weak equivalence principle in the context of GIE, but their approach does not involve internal degrees of freedom and therefore cannot address local position invariance. 
The present work fills these gaps by adapting the Zych-Brukner model of the quantum LPI to the analysis of GIE of quantum clocks.

\prlsection{Setup}
Our setup involves two  quantum mechanical particles whose internal state evolves in time and serves as a ``clock'', see \rFig{A}.
The two particles are placed closely to each other by a trapping potential so that they interact gravitationally.
Due to the mass-energy equivalence, the internal energy contributes to the mass of the particles, hence to the gravitational interaction.
The internal energy can in general be in a superposition of different values, in which case the particle behaves as if it is in a superposition of mass.
Thus, depending on the initial states and the internal Hamiltonians, entanglement will be generated between the two particles due to the gravitational interaction.
Our aim is to investigate how the possible violation of quantum local position invariance manifests in the behavior of this entanglement. 

We focus on the low-energy regime in which the gravitational interaction is described by the Newtonian potential.
For the simplicity of analysis, we assume that the two particles are identical. 
Thus, the internal states of the particles are represented by the same Hilbert spaces and Hamiltonians. 
To avoid complex analysis of the exchange interaction, we also assume that the overlap of the wave functions is negligible.


\prlsection{Adaptation of the Zych-Brukner model}
The Zych-Brukner model \cite{zych2018quantum,zych2017quantum} considers a single quantum particle in a classical gravitational field. 
If the equivalence principle holds, a particle's internal energy contributes equally to its rest mass, inertial mass and gravitational mass. 
In the non-relativistic limit, this gives the Hamiltonian $\hat{H}=\hat{H}_{\rm int}+c^2\hat{P}^2/2\hat{H}_{\rm int}+\hat{H}_{\rm int}\phi(\hat{Q})$, where $\hat{H}_{\rm int}$ is the internal Hamiltonian, $\hat{Q}, \hat{P}$ are the canonical position and momentum operators, and $\phi$ is the gravitational potential.
The three occurrences of $\hat{H}_{\rm int}$ correspond to the three types of mass described above, respectively. 
The quantum equivalence principle can then be tested by allowing these three Hamiltonians to be independent operators, $\hat{H}_{\rm rest}$, $\hat{H}_{\rm iner}$ and $\hat{H}_{\rm grav}$.
Thus, the test Hamiltonian is given by
\alg{
\hat{H}_{\rm ZB}=\hat{H}_{\rm rest}+\frac{c^2\hat{P}^2}{2\hat{H}_{\rm iner}}+\hat{H}_{\rm grav}\phi(\hat{Q}).
\laeq{ApHamiZB}
}
The quantum equivalence principle then asserts that the three Hamiltonian operators are equal.
In particular, the quantum LPI is represented as $\hat{H}_{\rm rest}=\hat{H}_{\rm grav}$.
The quantum LPI is violated either when the two Hamiltonians do not commute, or when the eigenvalues of them are not equal, or both. 
The former can be regarded as a ``genuinely quantum'' violation with no classical analog, while the latter can be viewed as a ``classical-like'' one \cite{zych2018quantum}.

We adapt the Zych-Brukner Hamiltonian, Eq.~\req{ApHamiZB}, to the setup described above. 
We consider two quantum clock particles trapped at fixed positions by an external trapping potential that acts only on their center-of-mass degrees of freedom.
The two-particle Hamiltonian will be obtained by summing up the one-particle Hamiltonian \req{ApHamiZB} for the two.
Since the particles are localized, their kinetic energy is negligible. 
The gravitational potential $\phi$ for one of the particles is the Newtonian potential sourced by the other.
While the passive mass (i.e. mass responsible for feeling gravity) and the active mass (i.e. mass responsible for sourcing gravity) are distinct notions \cite{kreuzer1968experimental,bartlett1986equivalence,singh2023equivalence,giulini2026if}, and a quantum model where they may differ can also be considered \cite{fragkos2025probing}, we here assume that the two masses are represented by a single Hamiltonian.
Thus, with $V_{A,B}$ denoting the trapping potential, the total Hamiltonian is given as
\alg{
\hat{H}_{\rm tot}=&\hat{H}_{\rm rest}^{A}+\hat{H}_{\rm rest}^{B}-\frac{G\hat{H}_{\rm grav}^{A}\otm\hat{H}_{\rm grav}^{B}}{c^4|\hat{Q}^{A}-\hat{Q}^{B}|}\nn\\
&\quad\quad+V_A(\hat{Q}^{A})+V_B(\hat{Q}^{B}),
\laeq{hamiltonian}
}
where $G$ is the Newtonian constant of gravitation and $c$ is the speed of light.
We treat Eq.~\req{hamiltonian} as an effective low-energy Hamiltonian that incorporates the possible violation of quantum LPI.
When the quantum LPI holds, i.e., $\hat{H}_{\rm rest}=\hat{H}_{\rm grav}$, and when the trapping potentials are dropped, the above Hamiltonian reduces to the one derived from quantum field theory \cite{castro2017entanglement,anastopoulos2014problems}.

\prlsection{Reduced dynamics of the internal states}
Suppose that the two-clock system is initialized in a state $\ket{\psi_{\rm ini}}$ and the external state is described by a wave function $\varphi_A(x)\varphi_B(y)$, where $x,y$ denote the position of the two particles.
The joint state is given by
$
|\Psi_{\rm ini}\rangle
=
\iint dxdy\:\varphi_A(x)\varphi_B(y)\ket{x}\ket{y}\ket{\psi_{\rm ini}}
$,
which will evolve in time under the Hamiltonian \req{hamiltonian}.
It is convenient to introduce a clock state $\ket{\psi_\ell(t)}=\hat{U}_\ell(t)\ket{\psi_{\rm ini}}\;(\ell>0)$, where
\alg{
\hat{U}_\ell(t)
=
\exp\left\{\frac{t}{i\hbar}\left(\hat{H}_{\rm rest}^{A}+\hat{H}_{\rm rest}^{B}-\frac{G\hat{H}_{\rm grav}^{A}\otm\hat{H}_{\rm grav}^{B}}{c^4\ell}\right)\right\}.
\laeq{psicr}
}
The state at time $t$ is then given as
$
|\Psi(t)\rangle
=
\iint dxdy\varphi_A(x)\varphi_B(y)e^{-i\theta(x,y)t}\ket{x}\ket{y}\ket{\psi_{|x-y|}(t)}
$,
where $\theta(x,y)=(V_A(x)+V_B(y))/\hbar$.
Since kinetic energy is neglected, the positions of the two particles do not change over time, so the spatial distribution $\varrho(\ell)$ remains unchanged.
The reduced dynamics of the two-clock state is obtained by tracing out the external degrees of freedom.
It reads
$
\rho(t)=\int d\ell\:\varrho(\ell)\proj{\psi_\ell(t)}
$,
where $\varrho(\ell)$ is the probability mass function defined by $\varrho(\ell)=\iint dxdy|\varphi_A(x)\varphi_B(y)|^2\delta(|x-y|-\ell)$.
We assume that the spatial spread of the wave function of each particle is modeled by a Gaussian distribution of width $w$, such as in the ground state of the harmonic potential,
which leads to $\varrho(\ell)=\frac{1}{\sqrt{2\pi w^2}}\exp(-(\ell-\ell_0)^2/2w^2)$, where $\ell_0(\gg w)$ is the average distance of the two particles.
We will be interested in evaluating the entanglement in the clock state $\rho(t)$.


\prlsection{Scheme 1: Entanglement generation from genuinely quantum LPI violation}
We first consider a scheme in which the possible violation of ``genuinely quantum'' LPI, namely, noncommutativity of $\hat{H}_{\rm rest}$ and $\hat{H}_{\rm grav}$, manifests in entanglement generation between the two clocks.
The initial clock state is chosen to be one of the eigenstates of the rest Hamiltonian.
We show, as the following theorem, that entanglement will be generated between the two clocks if and only if the two Hamiltonians do not commute.
Thus, generation of entanglement is a signature of the violation of genuinely quantum LPI.\\
{\bf Theorem:} 
Entanglement will be generated only if genuinely quantum LPI is violated, i.e., $[\hat{H}_{\rm rest},\hat{H}_{\rm grav}]\neq0$.
Conversely, if genuinely quantum LPI is violated, there exists an eigenstate $\ket{k}$ of the rest Hamiltonian such that, with $\ket{\psi_{\rm ini}}=\ket{k}\ket{k}$ and for sufficiently small $w$, the clock state at some time is entangled.\\
{\bf Proof:} 
The ``only if'' part is straightforward because the eigenstates of the rest Hamiltonian are invariant under the unitary evolution \req{psicr} if the two Hamiltonians commute.
To prove the ``if'' part, note that, in the limit of $w\rightarrow0$, the time evolution of the clock state is simply described by the unitary operator \req{psicr} for $\ell=\ell_0$. In this case, entanglement will be generated if $[\hat{H}_{\rm rest},\hat{H}_{\rm grav}]\neq0$. 
The statement follows by noting that the two-clock density operator continuously depends on $w$ and that the set of the non-entangled density operators is closed \cite{watrous2018theory}.
\QED

Let us consider an example of a two-level system.
The two Hamiltonians are expanded as $\hat{H}_{\rm rest}=E_g\proj{g}+E_e\proj{e}$ and $\hat{H}_{\rm grav}=\tilde{E}_g\proj{\tilde{g}}+\tilde{E}_e\proj{\tilde{e}}$, respectively.
The eigenvectors of the rest mass Hamiltonian are expanded by those of the gravitational mass Hamiltonian as
$
\ket{g}=\cos{\theta}\ket{\tilde{g}}+\sin{\theta}\ket{\tilde{e}},
\ket{e}=-\sin{\theta}\ket{\tilde{g}}+\cos{\theta}\ket{\tilde{e}}
$.
The noncommutativity of the two Hamiltonians is quantified as $\|[\hat{H}_{\rm rest},\hat{H}_{\rm grav}]\|_1={\rm Tr}|[\hat{H}_{\rm rest},\hat{H}_{\rm grav}]|=(E_e-E_g)(\tilde{E}_e-\tilde{E}_g)\sin{2\theta}$.
The above theorem states that clock entanglement will be generated at some time $t$ and for sufficiently small $w$ if and only if $\theta\neq m\pi/2\:(m\in\mathbb{N})$.


\prlsection{Scheme 2: Mismatch of entanglement frequency from classical-like LPI violation}
We next consider a scheme to investigate ``classical-like'' violation of LPI.
We assume that the ``genuinely quantum'' LPI is satisfied, i.e.,  the rest mass Hamiltonian and the gravitational mass Hamiltonian commute with each other.
Thus, the two Hamiltonians are simultaneously diagonalized as $\hat{H}_{\rm rest}=\sum_{k=0}^NE_k\proj{k}$ and $\hat{H}_{\rm grav}=\sum_{k=0}^N\tilde{E}_k\proj{k}$, where $\ket{k}$ is the $k$-th eigenvector of the Hamiltonians, and $E_k,\tilde{E}_k$ are the corresponding energy eigenvalues.
The LPI is then equivalent to $E_k=\tilde{E}_k\:(k=0,\cdots,N)$.
For the simplicity of analysis, we here focus only on the ground state $\ket{g}$ and the first excited state $\ket{e}$ of the rest Hamiltonian.
Nevertheless, the following argument applies to any pair of energy eigenstates with different eigenvalues.
The LPI implies that the energy gaps $\Delta{E}={E}_e-{E}_g$ and $\Delta\tilde{E}=\tilde{E}_e-\tilde{E}_g$ should be equal.
The violation of LPI can be quantified by deviation of the parameter $\eta:=\Delta\tilde{E}/\Delta{E}$ from unity.

In this scheme, the initial states of the clocks are chosen to be an equal superposition of the two energy eigenstates, that is, $\ket{\psi_{\rm ini}}=(\ket{g}+\ket{e})(\ket{g}+\ket{e})/2$. 
The expression of the state $\psi_\ell(t)$ is directly obtained from Eq.~\req{psicr},
in particular its one-clock reduced density matrix is calculated to be
$
\psi_\ell^A(t)
=
\frac{1}{2}(\proj{g}+\proj{e})
+\frac{1}{2}\cos{(\ell_0\tilde{\omega}_et/\ell)}(e^{i(\Omega-\ell_0\tilde{\omega}_r/\ell)t}\outpro{g}{e}+e^{-i(\Omega-\ell_0\tilde{\omega}_r/\ell)t}\outpro{e}{g})
$,
where the frequencies $\Omega$, $\tilde{\omega}_e$ and $\tilde{\omega}_r$ are given by
\alg{
\Omega=\frac{\Delta E}{\hbar},\quad
\tilde{\omega}_e=\frac{G\Delta \tilde{E}^2}{2\hbar c^4\ell_0},\quad
\tilde{\omega}_r=\frac{G\tilde{M}\Delta\tilde{E}}{\hbar c^2\ell_0},
\laeq{entfreq}
}
and $\tilde{M}=(\tilde{E}_g+\tilde{E}_e)/2c^2$.
The entanglement entropy of the two-clock state $\psi_\ell(t)$ is calculated to be
$
E(\psi_\ell(t))=
h\left(\frac{1}{2}+\frac{1}{2}\cos{(\ell_0\tilde{\omega}_et/\ell)}\right)
$,
where $h$ is the binary entropy defined by $h(x)=-x\log{x}-(1-x)\log(1-x)$.

It is instructive to note that $\tilde\omega_r$ is the gravitational redshift on clock A due to the effective gravitational potential generated by the average mass of clock B.
It thus has a direct analog in single-clock tests with a classical gravitational source \cite{zych2018quantum,zych2017quantum}.
In contrast, $\tilde{\omega}_e\propto\Delta\tilde{E}^2$ originates from the operator product $\hat{H}_{\rm grav}^A\otimes\hat{H}_{\rm grav}^B$, vanishes if either mass is classical, and coincides with the frequency at which the entanglement of the two clocks oscillates.
Thus, it is a signature specific to a quantum source of gravity.

The frequency $\tilde{\omega}_e$ can be evaluated by performing measurements of the observables $X=\outpro{e}{g}+\outpro{g}{e}$ and $Y= i\outpro{e}{g}-i\outpro{g}{e}$ on one of the clocks at each time, e.g., by Ramsey interferometry. 
Indeed, as shown in \rApp{scheme2gauss}, the expectation values of these observables with respect to the average density matrix $\rho^A(t)=\int d\ell\varrho(\ell)\psi_\ell^A(t)$ satisfy 
$
\langle X\rangle^2+\langle Y\rangle^2
=
\frac{1}{4}\left[\Gamma_1(t)+\Gamma_2(t)\cos{2\tilde{\omega}_et}\right]
$,
where $\Gamma_{1,2}$ are monotonically decreasing functions that describe the decoherence of the clock state due to the spatial spread of the wave functions.

The LPI parameter $\eta$ is represented in terms of the distance of the two particles $\ell_0$, the entanglement frequency $\tilde{\omega}_e$ given by Eq.~\req{entfreq}, and the proper frequency of each clock, as
\alg{
\eta^2
=
\frac{2c^4}{\hbar G}\cdot\frac{\ell_0\tilde{\omega}_e}{\Omega^2}.
\laeq{scheme2QEP}
} 
Any deviation of the R.H.S. from unity is thus a signature of the violation of LPI.
Note that $\ell_0$ is a controllable parameter, and that both $\tilde{\omega}_e$ and $\Omega$ can, in principle, be measured by monitoring the time evolution of entanglement in the presence of gravitational interaction and that of the internal clock state without gravity.


\prlsection{Complementarity of the two schemes}
\lapp{}
The two schemes presented above can be realized in a single setting simply by changing the initial state of the clocks.
In Scheme 1, the initial state is chosen to be one of the eigenstates of the rest Hamiltonian, whereas in Scheme 2, the initial state is a superposition of the eigenstates thereof.

The two schemes may draw different conclusions about the possible violation of quantum LPI.
In Scheme 1, generation of entanglement immediately implies the violation of LPI.
It is a signature of the genuinely quantum violation of LPI, that is, noncommutativity of the rest mass Hamiltonian and the gravitational mass Hamiltonian.
In Scheme 2, observation of entanglement alone does not imply the violation of LPI.
Instead, the violation of LPI manifests in the deviation of the frequency of entanglement oscillation from the expected one.
The LPI violation that could be observed in Scheme 2 is the difference of the eigenvalues of the two Hamiltonians.
Our approach thus provides a unified framework to address two different but complementary aspects of the quantum LPI, in a single setup with different choices of the initial states.


\prlsection{Related work}
Our work is most directly related to the original proposal of Zych and Brukner \cite{zych2018quantum,zych2017quantum}, which established the distinction between genuinely quantum (eigenvector) and classical-like (eigenvalue) violations of LPI, and proposed an interferometric test based on a single quantum clock within a classical gravitational field. 
The present work extends this distinction to the setting in which the source of gravity is itself a quantum system in superposition, using gravity-induced entanglement between two clocks in place of interference of a single clock.
The idea of extending the Zych-Brukner model to the case of quantum background spacetime, and observing its possible violation through GIE, was originally introduced in the author's earlier work \cite{wakakuwa2025detectability}.
While it focused specifically on post-Newtonian effects such as frame dragging, the present work develops this idea in the more basic setting of the Newtonian GIE.

Ref.~\cite{bose2023entanglement} investigated a related protocol to witness possible violation of the quantum equivalence principle in the context of gravity-induced entanglement.
It is based on an adaptation of the Bose-Marletto-Vedral protocol~\cite{bose2017spin,marletto2017gravitationally} to probe the weak equivalence principle via spatial superpositions of gravitationally interacting particles. They showed that the amount of generated entanglement differs depending on whether the weak equivalence principle is violated.

We highlight two key differences between the approach of \cite{bose2023entanglement} and ours. First, the two approaches address different aspects of the equivalence principle. Ref.~\cite{bose2023entanglement} focuses on the weak equivalence principle (the universality of free fall), formulating QEP violation through the ratio of inertial to gravitational mass treated as c-numbers. 
Our approach, on the other hand, addresses local position invariance (LPI), formulated as the equality between the rest mass and gravitational mass Hamiltonians \cite{zych2018quantum}. 
By treating these masses as operators rather than c-numbers, our formulation allows us to distinguish two qualitatively different types of LPI violation, that is, differing eigenvectors versus differing eigenvalues of the two Hamiltonians.
These two manifest in distinct signatures, and captures genuinely quantum mechanical aspects of QEP violation with no classical analog.

Second, the two approaches differ in how QEP violation manifests in observable quantities. 
In Ref.~\cite{bose2023entanglement}, the violation appears as a change in the entanglement generation rate, and this must be compared with the rate predicted from the particle separation and the applied magnetic force. 
In our Scheme 2, the violation appears analogously, as a mismatch between the observed frequency of entanglement oscillation and the one expected from the clocks' proper energy gap and spatial separation. 
In Scheme 1, by contrast, the manifestation is qualitatively different: the mere presence of entanglement itself is a signature of QEP violation, independent of any comparison to other quantities.


\prlsection{Conclusion}
We have proposed a framework for analyzing the quantum local position invariance (LPI) in the context of gravitationally induced entanglement (GIE) of quantum clocks. 
By extending the formulation of LPI by Zych and Brukner to the two-clock setting, we have analyzed two complementary schemes differing only in the initial clock state and addressed two distinct types of quantum LPI violations.
Our results suggest that GIE of quantum clocks provides a natural framework to investigate the geometric interpretation of gravity in the quantum regime, where the validity of the equivalence principle can no longer be taken for granted.

We note that our framework is not contingent on the resolution of the ongoing debate on whether the detection of GIE implies the quantum nature of gravity (see e.g.~recent literature \cite{AzizHowl2025,DiBiagio2025ClassicalEntangle,MarlettoVedral2025ClassicalGravity,mitrakos2026does,trillo2025diosi,ludescher2026gravity,weller2025there,angeli2025entanglement,marletto2025classical,aspelmeyer2026quantum,feng2026collapse,gundhi2026can,schneider2025demonstration,lin2026can,diosi2025no,boulle2025subsystems,xue2026aziz,marletto2025quantum} and the references therein). 
Even if this inference were found to be invalid, our framework would retain its value as a means for investigating the quantum equivalence principle in a
regime where the source of gravity is treated quantum mechanically, providing an approach that is not accessible via classical spacetime experiments.

Several directions remain for future work. 
As with other proposals for gravity-induced entanglement, the weakness of the gravitational coupling poses a formidable experimental challenge. A quantitative analysis of feasibility, identifying dominant noise sources and estimating required coherence times relative to the parameters identified here, remains an important direction for future work.
It would also be interesting to extend our framework to incorporate the distinction between active and passive gravitational mass, which are identified in the present model but represent independent aspects of the equivalence principle \cite{fragkos2025probing}.
Finally, a more systematic study of the relationship between QEP violation and the breakdown of the geometric interpretation of gravity in the quantum regime would help clarify the broader implications of the signatures of LPI violation identified in this work.

\prlsection{Acknowledgments}
This work is supported by MEXT Quantum Leap Flagship Program (MEXT QLEAP), Grant No.~JPMXS0120319794, and by JSPS KAKENHI, Grant No.~JP26K07065.

\bibliography{bibliography.bib}

%
%

\appendix

\begin{widetext}

\section{Evaluation of entanglement in Scheme 2}
\lapp{scheme2gauss}

Suppose that the probability mass function $\varrho(\ell)$ is represented as the Gaussian distribution centered around a fixed radius $\ell_0$ with the width $w$:
\alg{
\varrho(\ell)=\frac{1}{\sqrt{2\pi w^2}}\exp\left(-\frac{(\ell-\ell_0)^2}{2w^2}\right).
}
With $\xi$ denoting $\ell-\ell_0$, it can be rewritten as
\alg{
\varrho(\xi)=\frac{1}{\sqrt{2\pi w^2}}\exp\left(-\frac{\xi^2}{2w^2}\right).
}
Suppose that the width $w$ of the wave function is sufficiently small compared to the average distance $\ell_0$ of the two particles.
Then, in the first order of $\xi$, we have
\alg{
\ell=\ell_0+\xi,
\quad
\frac{\ell_0}{\ell}
\approx
1-\frac{\xi}{\ell_0}.
}
Let $\omega\in\mbb{R}_+$ be some frequency.
By the Gaussian integral, we have
\alg{
\int_{-\infty}^\infty\varrho(\ell)\exp\left( i\frac{\ell_0}{\ell}\omega t\right)d\ell
&\approx
\int_{-\infty}^\infty\varrho(\xi)\exp\left( i\left(1-\frac{\xi}{\ell_0}\right)\omega t\right)d\xi\\
&=
\frac{1}{\sqrt{2\pi w^2}}\int_{-\infty}^\infty\exp\left(-\frac{\xi^2}{2w^2}+ i\left(1-\frac{\xi}{\ell_0}\right)\omega t\right)d\xi\\
&=
\exp\left(-\frac{w^2\omega^2t^2}{2\ell_0^2}+ i\omega t\right)\\
&=
\exp\left(-\frac{\gamma_\omega^2t^2}{2}+ i\omega t\right),
\laeq{muk}
}
where $\gamma_\omega=w\omega/\ell_0$.

For each $\ell$, the one-clock reduced density operator is given by
\alg{
\psi_\ell^A(t)
&=
\frac{1}{2}(\proj{g}+\proj{e})
+\frac{1}{2}\cos{\left(\frac{\ell_0\tilde{\omega}_et}{\ell}\right)}\left(e^{i\left(\Omega-\frac{\ell_0\tilde{\omega}_r}{\ell}\right)t}\outpro{g}{e}+e^{-i\left(\Omega-\frac{\ell_0\tilde{\omega}_r}{\ell}\right)t}\outpro{e}{g}\right)\\
&=\frac{1}{2}(\proj{g}+\proj{e})
+\frac{1}{4}\left[\left(e^{i\left(\Omega-\frac{\ell_0(\tilde{\omega}_r+\tilde{\omega}_e)}{\ell}\right)t}+e^{i\left(\Omega-\frac{\ell_0(\tilde{\omega}_r-\tilde{\omega}_e)}{\ell}\right)t}\right)\outpro{g}{e}+H.c.\right].
\laeq{psir1t}
}
Applying \req{muk}, the averaged one-clock density operator is given by
\alg{
\rho^A(t)
=
\int\varrho(\ell)\psi_\ell^A(t)
=
\frac{1}{2}(\proj{g}+\proj{e})
+\frac{1}{4}\left[e^{i\Omega t}\left(e^{-\frac{\tilde{\gamma}_+^2t^2}{2}-i(\tilde{\omega}_r+\tilde{\omega}_e) t}+e^{-\frac{\tilde{\gamma}_-^2t^2}{2}-i(\tilde{\omega}_r-\tilde{\omega}_e) t}\right)\outpro{g}{e}+H.c.\right],
}
where $\tilde{\gamma}_\pm=w(\tilde{\omega}_r\pm\tilde{\omega}_e)/\ell_0$.
Thus, we have
\alg{
\langle X\rangle_{\rho^A(t)}
&=\frac{1}{2}\left[e^{-\frac{\tilde{\gamma}_+^2t^2}{2}}\cos{(\Omega-\tilde{\omega}_r-\tilde{\omega}_e) t}+e^{-\frac{\tilde{\gamma}_-^2t^2}{2}}\cos{(\Omega-\tilde{\omega}_r+\tilde{\omega}_e) t}\right],
\\
\langle Y\rangle_{\rho^A(t)}
&=-\frac{1}{2}\left[e^{-\frac{\tilde{\gamma}_+^2t^2}{2}}\sin{(\Omega-\tilde{\omega}_r-\tilde{\omega}_e) t}+e^{-\frac{\tilde{\gamma}_-^2t^2}{2}}\sin{(\Omega-\tilde{\omega}_r+\tilde{\omega}_e) t}\right],
}
which yields
\alg{
\langle X\rangle_{\rho^A(t)}^2+\langle Y\rangle_{\rho^A(t)}^2
=
\frac{1}{4}\left[e^{-\tilde{\gamma}_+^2t^2}+e^{-\tilde{\gamma}_-^2t^2}+2e^{-\frac{(\tilde{\gamma}_+^2+\tilde{\gamma}_-^2)t^2}{2}}\cos{2\tilde{\omega}_et}\right].
}

\end{widetext}

\end{document}